# Embedded metal nanopatterns for near-field scattering-enhanced optical absorption

Fan Ye, Michael J. Burns, and Michael J. Naughton*
Department of Physics, Boston College, 140 Commonwealth Avenue, Chestnut Hill, MA 02467, USA

Abstract

Simulations of metal nanopatterns embedded in a thin photovoltaic absorber show significantly enhanced absorbance within the semiconductor, with a more than 300% increase for λ = 800 nm. Integrating with AM1.5 solar irradiation, this yields a 70% increase in simulated short circuit current density in a 60 nm amorphous silicon film. Embedding such metal patterns inside an absorber maximally utilizes enhanced electric fields that result from intense, spatially organized, near-field scattering in the vicinity of the pattern. Appropriately configured (*i.e.* with a thin insulating coating), this optical metamedium architecture may be useful for increasing photovoltaic efficiency in thin film solar cells, including offering prospects for realistic ultrathin hot electron cells.

Traditional solar cell architectures are principally planar [1], with contact and photovoltaic (PV) junction films stacked upon each other in a sandwich structure. In this arrangement, normally-incident light travels parallel to the direction that photogenerated carriers need to diffuse in order to reach contacts for useful current extraction. As a result, light absorption and charge diffusion compete due to disparities between optical and electronic length scales. An optimized solar cell would thus be simultaneously thick enough to maximize optical absorbance and thin enough to maximize charge extraction. Various designs have been proposed to circumvent this "thick-thin" challenge. One orthogonalizes and thus decouples the optics and electronics by forming radial PV junctions perpendicular to incident light, such as nanocoax [2, 3] and nanowire [4, 5] solar cells. Here, thinner PV films can be employed to improve charge extraction while using the structure itself to improve light trapping.

Another design maintains the planar structure and re-lies on extending the optical path length within the PV layers by scattering incident and/or internally-reflected light at oblique angles. This allows PV films to be made thinner, as light traveling at off-angles increases its path length. Examples of this are planar cells with surface texturing, and/or inclusion of dielectric or metal particles as scatterers. Dielectric or metal [6–8] nanoparticles or patterns [9–14] placed atop an absorber layer can be used to increase off-angle forward (e.g. dipole) scattering as well as to attempt to couple far-field light into waveguide modes of the absorber, thereby prolonging the optical path length. When applied to the back reflector [15–19], the intent is either to recycle unabsorbed photons by creating multiple passes through the absorber through specular reflection, or again to couple into lateral waveguides. With subwavelength-sized and/or spaced metal particles, further light manipulation can be achieved by exciting localized or propagating surface plasmon modes [20].

These top and bottom surface strategies, however, only capitalize on *E*-field enhancement at the one side that abuts the PV medium. The top contact is usually a transparent conductive oxide, while a typical back contact is a highly optically-reflecting metal. While such materials absorb only small fractions of incident light, their proximity to any of the above strategies means that absorption is not maximized. However, positioning strong scatterers directly inside (fully surrounded by) an absorber constitutes a potentially improved route to enhancing optical absorption in thin film PV, by allowing one to harvest all internally scattered light, and thus increase the number of photogenerated electron-hole pairs. Issues of light interacting with particles embedded in absorbing and nonabsorbing media were considered in the late 1800's by Stokes, Tyndall, Lorenz and Strutt (Rayleigh), among others, who generally considered scattering from spherical particles much smaller than the wavelength of light. Mie [21] later derived solutions to Maxwell's equations for scattering from spheres of size comparable to and larger than the wavelength, but only in the presence of nonabsorbing media. More recently, modified Mie solutions have been derived for scattering in absorbing media, initially using far-field [22–25] and later, near-field [26, 27] analytical analyses.

Recently, we simulated optical absorbance in thin film amorphous silicon (*a*-Si) integrated not with spherical met-al particles above or below the absorber, but with embedded metal nanopatterns (EMN) positioned directly in the film [28], and these yielded increases in absorbance relative to a control *a*-Si film without such integration. We then fabricated a test sample with the cross pattern, and found a 40% increase in absorbance over the control. The dimensions of this cross were only marginally subwavelength, but the results motivated us to investigate in more detail the effect of embedding metal nanopatterns in PV absorbers.

Here, we present detailed results of computer simulations on optical absorption in ultrathin *a*-Si integrated with a subwavelength-dimensioned EMN, forming an optical metamedium. The simulations show enhanced absorption within the *a*-Si (*i.e.* not in the embedded metal), especially at long wavelengths. We interpret the effect as due to enhanced evanescent near-field scattering from the EMN metamedium, and not from waveguide or plasmonic resonance effects. They also provide insight into how one might tailor this effect with the architecture of the EMN metamedium. The EMN scatterers discussed herein, while all subwavelength (10's to 100's of nanometers) straddle the Strutt and Mie size domains and, moreover, are non-spherical, making exact analytical analysis nontrivial. Nonetheless, the observed phenomena can be rather accurately simulated with modern computational techniques that solve Maxwell's equations with appropriate boundary conditions [27-32]. The structured aspect of the nonspherical EMN discussed here reflects the fact that it functions as a quasi-broadband optical metamaterial, with many of the novel radiation-manipulating capabilities of RF, THz and IR metamaterial systems. Employing *a*-Si films of the ultrathin nature discussed here in solar cells has several benefits beyond enhanced absorption: reduced Staebler-Wronski [33,34] light-induced degradation, improved junction quality, and the potential for hot electron effects [35], all of which can result in power conversion efficiencies well above existing records.

A sketch of the EMN cross scheme discussed in this paper is shown in Fig. 1, with side and plan views of a simplified thin film solar cell-like design: a semiconductor (*a*-Si) sandwiched by a transparent top (fluorine-doped tin oxide, FTO) and reflective bottom (Ag) contact. A *bona fide a*-Si solar cell will include *p*- and *n*-doped a-Si films above and below (or vice versa) the undoped (intrinsic *i*-layer) *a*-Si shown, as well as a passivation layer (*e.g.*, ZnO) between the silicon and silver. However, these details only marginally affect the optical performance of a so-designed cell, and are thus neglected for simplicity in the simulations presented herein. Visible light injected from the FTO side is either reflected or absorbed by the 3-layer stack, as the Ag back reflector is optically thick, precluding transmission. As such, the back reflector thickness is not germane to the problem, and for simulations, relevant parameters are the thicknesses of the EMN ($h_{EMN}$), *a*-Si ($h_{Si}$) and FTO layers, and the EMN structure and position. As shown in Fig. 1, an EMN is positioned a distance *d* from the top surface of the *a*-Si layer (measured downward to the top of the EMN film). By varying *d*, we can examine all possible configurations for which such a pattern can be integrated into an absorber film, with $d \leq -h_{EMN}$ representing top-patterning or above, $-h_{EMN} < d \leq 0$ partially embedded, $0 < d < h_{Si} - h_{EMN}$ fully embedded, and $d \geq h_{Si} - h_{EMN}$ bottom-patterning (*i.e.* on or in the Ag back reflector). As a result, we are able to address in detail the issue of the optimum depth to position a particular subwavelength-dimensioned, cross-shaped EMN, as well as the physical origin of the optical absorbance enhancement.

The simulated cross EMN is comprised of an array of $h_{EMN}$ = 20 nm thick Ag crosses with 100 × 300 nm² segments in a 400 × 400 nm² unit cell (four cells shown in Fig. 1). Larger, microscale versions of this cross structure have been employed as components of metamaterials with a geometrical resonant absorption in the mid-infrared when positioned on a dielectric spacer that is coupled to a metal plane [36]. For the size used here, such resonant response is significantly modified and outside the visible wavelength regime, with a combination of a large blue shift due to the reduced size, a small red shift due to embedding, and strong damping and broadening due to embedding in the absorbing *a*-Si medium. For our simulated structure, we chose 50 nm thick FTO and $h_{Si}$ = 60 nm thick silicon, and varied the EMN embedding depth *d* from –30 to +60 nm (*i.e.* fully within the FTO to fully within the back Ag reflector). The simulations consisted of placing periodic boundary conditions on the unit cell in the plane of the EMN, and simulating the response to a normally-incident, linearly polarized plane wave ($E_x$, see Fig. 2). The $0^{th}$ order reflectance *R* and transmittance *T* of the incident wave normal to the surface were simulated, yielding absorbance $A = 1 - R$ (since $T = 0$ for the totally reflective back contact). Simulations were performed using commercial, finite element analysis tools COMSOL Multiphysics and CST Microwave Studio in the frequency-domain, with portions of the simulations cross-checked between the two software packages.

Full dispersion relations from standard literature sources were employed for all materials (*a*-Si [37], Ag [38–40], FTO [41]) in the simulations. Moreover, at the suggestion of Green and Pillai [42], we compared simulations using optical constants for silver from Refs. [38–40], finding negligible differences in the results.

Enhanced absorption within *a*-Si Figure 2(a) shows the resulting simulated absorbance within the *a*-Si layer, obtained by integrating the calculated, time-averaged power loss density $P_L$ over the *a*-Si volume, versus incident light wavelength, for different embedding depths *d*. The dashed line shows the simulated absorbance without an integrated EMN. Here,

$$P_L = \frac{\omega \varepsilon''}{2}|\vec{E}|^2 + \frac{\omega \mu''}{2}|\vec{H}|^2 \qquad (1)$$

is derived from Poynting's theorem [43], with ω the light frequency, $\varepsilon''$ and $\mu''$ the imaginary parts of the relative complex dielectric constant and permeability of the absorber, and *E* and *H* the local electric and magnetic fields, respectively. Since $\mu'' \approx 0$ for *a*-Si, the magnetic term does not contribute to absorption. It can be seen from the figure that when the Ag cross pattern is present but positioned above the *a*-Si layer (*d* = −30 nm, i.e. bottom surface of EMN lying 10 nm above the FTO/*a*-Si interface), the absorbance in *a*-Si is less than the bare *a*-Si, "no EMN" condition, across most of the 350 nm to 850 nm range investigated. This appears rational since Ag, which in the employed cross pattern covers 5/16 ~ 38% of the exposed surface, is known to be highly reflective in this frequency range. As the Ag pattern is brought closer to and then embedded into the *a*-Si layer (as *d* is increased from –30 nm to 0 nm), however, an overall increase in absorbance is ob-served throughout the spectrum, but especially at long wavelengths (λ > 600 nm). As the embedding depth is further increased, the total absorbance in *a*-Si does as well, until it reaches a maximum between *d* = 10 nm and 20 nm. Finally, continued increases in *d* (*e.g.*, +30 to +40 nm) ultimately suppress absorbance, again especially so at long wavelengths. The absolute absorbance in the

*a*-Si in the simulations in Fig. 2(a) reaches more than 80% for this optimum embed depth regime.

For added perspective, the simulations of Fig. 2(a) can be represented as a contour plot of *a*-Si absorbance versus wavelength, shown in Fig. 2(b), with a 0–1 linear color scale on the right. From this plot, one can discern an optimum embedding depth near $d$ = 15 nm for which a some-what broad, high absorption band develops. Significant below-gap (*i.e.*, above $\lambda = hc/eE_g \sim 700$ nm, where $E_g \sim 1.7$ eV is the nominal band gap of *a*-Si) absorption can be seen for EMN depths near the vertical middle of the structure, as opposed to those near the top FTO and bottom Ag surfaces. From these data in this structure, we can also explicitly ex-tract the depth dependence of the absorbance $A(d)$ within the *a*-Si volume, for fixed wavelengths. We have selected the free-space wavelengths $\lambda$ = 500, 600, 700 and 800 nm for display, as indicated by arrows at the top of the contour plot. Normalized to the control, "no EMN" absorbance $A_o$, we plot the resulting optical absorbance enhancement factor $A(d)/A_o$ versus $d$ in Fig. 2(c). This is the main result of this paper: embedding a metal nanopattern inside an optical absorber can enhance, by significant amounts, the optical absorbance of the surrounding semiconductor medium. For the example structure shown here, the enhancement is ~175% at $\lambda$ = 700 nm ($A/A_o > 2.5$), and ~325% at $\lambda$ = 800 nm. The detail shown in Fig. 2(c) provides further evidence for an optimum embedding depth in the range $d$ = 10 to 20 nm for the wavelength values depicted, but especially for the longer, near- and sub-gap wavelengths. Note again that the more conventional top ($d$ = −20 nm) and bottom ($d \geq$ +40 nm) pattern placements yield far less enhancement at most wavelengths, as compared to the embedded situations.

One can estimate the effect this EMN concept can have on a photovoltaic solar cell, within the assumptions that *p*- and *n*-doped layers on either side of the undoped *a*-Si film do not appreciably change the optical absorbance, and that this absorbance can be equated with external quantum efficiency. Using for the short circuit current density $J_{sc}$ =

$(e/hc)\int S(\lambda)A(\lambda)\lambda d\lambda$ where $e$, $h$ and $c$ are the elementary charge, Planck's constant and the speed of light, respectively, and $S(\lambda)$ is the power density spectrum for solar irradiation (AM1.5) [44], we can calculate $J_{sc}$ for each embedded depth $d$. We plot in Fig. 2(c) the ratios of these $J_{sc}$ values to that for the 'no EMN' control, calculated to be $J_{sco}$ = 12.4 mA/cm$^2$. Consistent with the simulated $A(\lambda)$, $J_{sc}$ is enhanced upon embedment, peaking near $d$ =15 nm with a 70% increase over the control ($J_{sc}$ > 21 mA/cm$^2$), a value 20% higher than the record number for single junction $a$-Si [45], achieved in more than 4 times thicker films. For a typical open circuit voltage of $V_{oc}$ = 0.88 V and fill factor of 0.7 for $a$-Si solar cells, this corresponds to a power conversion efficiency η of 13%, close to 30% higher than the state-of-the-art for single junction $a$-Si PV [46].

Insight into the physical origin of this metamedium-enabled increase in optical absorbance is obtained by examining the spatial dependence of the power loss density $P_L$, which again is a local measure of optical intensity or absorbance ($P_L \sim E^2 \sim A$) within the volume of the structure, extracted from the simulations. To this end, we show in Fig. 3 the simulated $P_L$ for a series of cross-sections and slices of the Ag cross EMN of Figs. 1 and 2, for normally-incident, $x$-polarized, $\lambda$ = 700 nm light at various embed-ding depths. The top panel on the left set of images shows $P_L(y,z)|_x$ for the situation without the Ag EMN (i.e. bare $a$-Si), as well as the orientations of light propagation $k_z$ and electric field polarization $E_x$. Successive panels below this correspond to $y$-$z$-plane cross-sections of $P_L$ at depths between $d$ = −20 and +40 nm, in 10 nm steps, for a cut (i.e. fixed $x$) through the middle of the crosses in the EMN array. All panels contain the same linear color scale shown, varying from 0 to 5×10−10 W/m$^3$, which corresponds to an optical density equivalent to ~1 sun (1 kW/m$^2$). These images serve to implicate the mechanism responsible for the optical absorbance enhancement: near-field scattering-enhanced concentrations of $P_L$ within the $a$-Si in the vicinity of the embedded nano-crosses, strongest in intensity above and between crosses for the $d$ = 10 and 20 nm EMN depths. Upon embedment, the scattered electric field concentrates in the *a*-Si, which has a much higher absorption coefficient than FTO or bulk Ag at optical frequencies.

The patterned aspect of the EMN is responsible for it functioning as a type of optical metamaterial, focusing scattered light in structure-defined volumes of the *a*-Si absorber. Such a structure has a multitude of length scales that, compared to simpler structures such as spheres or oblate spheroids, may lead to a broadband scattering response [47]. The role of plasmonics in the observed absorption enhancement is not clear, however. For a Ag/*a*-Si interface, a surface plasmon polariton resonance is expected to occur when the real parts of the dielectric constants of silver and silicon match, *i.e.* $\text{Re}[\varepsilon_{Ag}(\omega)] = -\text{Re}[\varepsilon_{a\text{-Si}}(\omega)]$. Using experimental dielectric responses from the literature [38-40], this occurs at $\omega_{SPP}$ = 3.022 PHz, corresponding to $\lambda_{SPP}$ = 623 nm [48]. Referring to Fig. 2(a), no distinct or resonance-like feature appears near or above this wavelength. The right column of images in Fig. 3 shows *x-y*-plane slices at various depths *z* of the simulated $P_L$ for one unit cell of this EMN array, at the *d* = 20 nm embed depth. Here, the spatial distribution of the near-field-enhanced scattering can be further appreciated, as well as the low absorption by the Ag EMN itself (*e.g.* the *z* = 30 nm slice through the midplane of the EMN). While not shown here, simulations using unpolarized light show the same field enhancements, but without the obvious in-plane anisotropy evident in Fig. 3. We have also simulated these effects for EMNs having a thin (5 nm) insulating coating, and found only a few per cent change in the absorbance in the *a*-Si (in fact, a 2% absolute increase for the *d* = 20 nm case). Along with the issues of PV junction quality, cell stability and hot electron effects mentioned above, this is an essential point with regard to potential utilization of the EMN concept in photovoltaics.

In conclusion, electromagnetic simulations show that a metamedium comprised of a subwavelength-sized metal nanopattern embedded in an optical absorber exhibits a spatially inhomogeneous electromagnetic response, with incident light intensely scattered, and to an extent focused, into localized regions within the absorber. This organized near-field scattering effect leads to strongly enhanced absorbance in these regions and, accounting for the whole sample volume, significant increases in short circuit current (+70%) and photovoltaic performance (+30%) over that of a control. The enhancement is particularly strong in the near infrared, more than 4 times that in the control at $\lambda = 800$ nm. The EMN-induced absorbance and $J_{sc}$ enhancements obtained here exceed the highest reported values from other simulations for nanoparticles on top [6–9], bottom [17, 18] or embedded [30, 31], or grids on top [10–13] or bottom [15, 16] of a PV film. While the subwavelength scale of the EMN presented in this report is essential, its precise shape is not considered to be optimal with respect to optical absorbance in *a*-Si, let alone other PV absorbers, such that future design optimization may result in even further enhancements. With regard to eventually technical and economical feasibility of EMN-based photovoltaics, we note that there exist several scalable processes for such integration, such as roll-to-roll stamp imprinting or nanosphere lithography. Finally, fabrication of insulated EMNs within a *p-i-n* junction *a*-Si thin film (or other) solar cell will be needed to determine the extent to which this absorbance indeed translates into increased photovoltaic efficiency.

**Acknowledgements**

This work was supported in part by the W. M. Keck Foundation.

**Figure Captions**

Figure 1. Sketch of embedded metal nanopattern (EMN) scheme, with cross-section of idealized absorber structure having integrated Ag (gray) EMN in *a*-Si (red), with cross EMN. The embedding depth *d* is indicated.

Figure 2. Simulated absorbance *A* within *a*-Si while tuning the embedding depth *d* of a Ag cross EMN for normally-incident, linearly polarized light (50 nm FTO, 60 nm *a*-Si and 20 nm EMN thicknesses). (a) *A* vs. free-space wavelength for various *d*, for EMN placement between on-the-top ($d \leq -20$ nm) and on-the-bottom ($d \geq +40$) contacts, showing strong near infrared enhancement. (b) Contour plot of absorbance data in (a) on linear 0–1 color scale, highlighting the optimum embed depth regime. (c) Left scale: Absorbance enhancement at fixed wavelengths vs. embed depth *d*, relative to an EMN-free control sample, showing strongest effects at long wavelengths (>300% at 800 nm). Right scale: Variation in calculated short circuit current density with *d*, relative to the control, showing maximum ~70% increase near *d* = 15 nm.

Figure 3. Simulated power loss density at $\lambda = 700$ nm linearly polarized ($E_x$) incidence for the Ag cross EMN, viewed in *y*-*z*-plane cross-sections cut through the middle of two unit cells, for various embed depths *d*. The light-sample coordinate system is indicated, and only the FTO and *a*-Si layers are shown, with the EMN indicated by its outline. The $0-5 \times 10^{-10}$ W/m$^3$ linear color scale is shown. It can be seen that $P_L (x,y,z)$ reaches maximal values between *d* = 10 and 20 nm, between and above individual crosses, respectively. On the right is a series of *x*-*y* slices of the *d*=20 nm depth $P_L$ at different *z*-positions, providing a separate perspective of the spatial distribution of electromagnetic absorption, which is primarily in the *a*-Si.

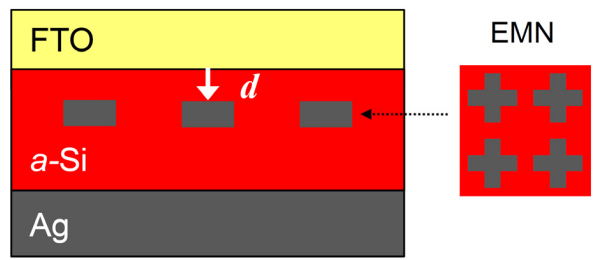

Figure 1.

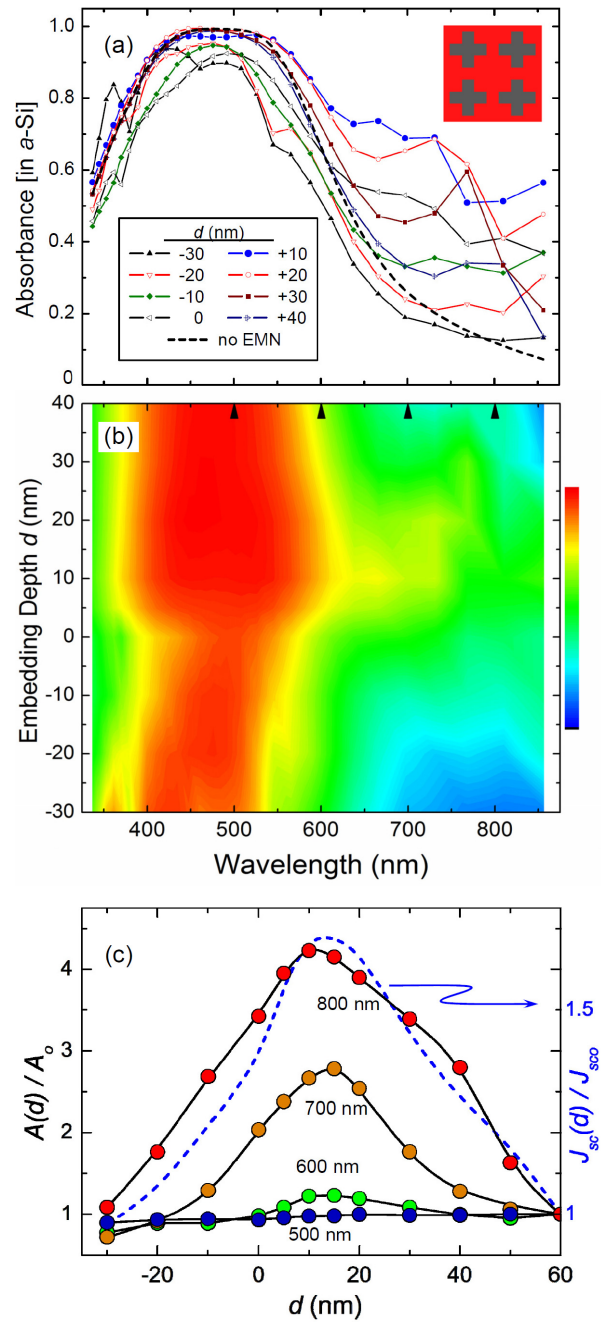

Figure 2.

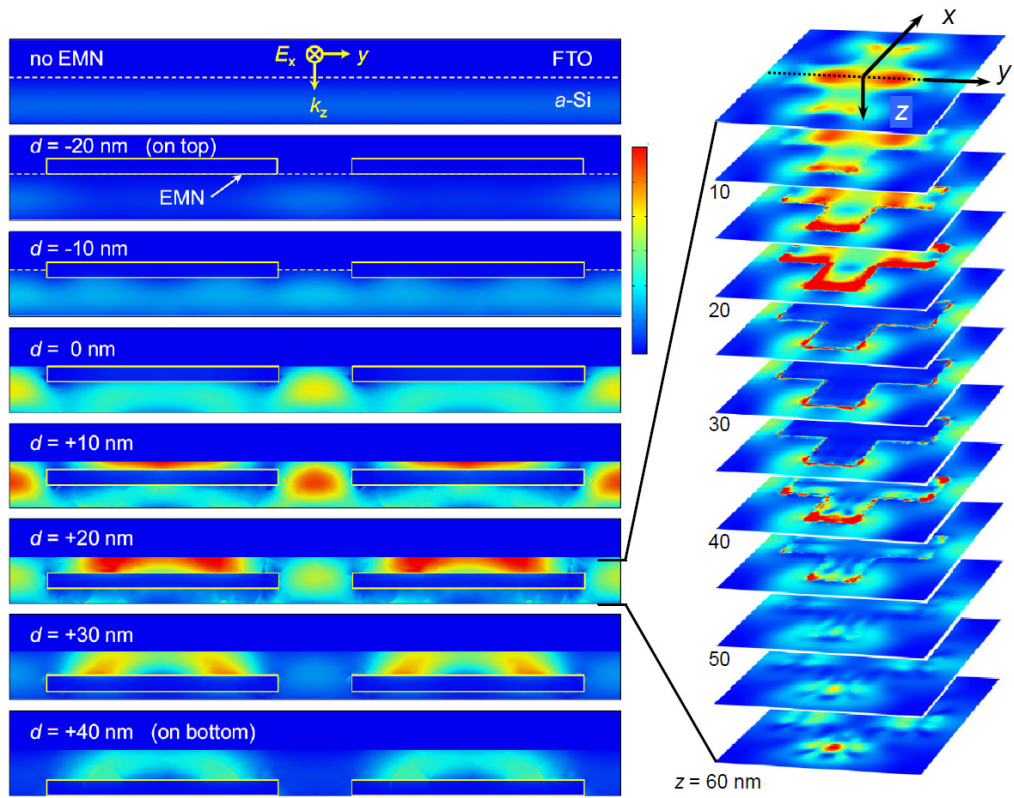

Figure 3.